\documentclass[aps,reprint,superscriptaddress,nofootinbib,twocolumn]{revtex4-2}
\pdfoutput=1
\usepackage[utf8]{inputenc}\usepackage[T1]{fontenc}
\usepackage{amsmath}
\usepackage{amssymb}
\usepackage{graphicx}
\usepackage{comment}
\usepackage{slashed}
\usepackage{cases}
\usepackage[toc,page]{appendix}
\usepackage{empheq}
\usepackage[normalem]{ulem} % to use sout
\usepackage{xcolor} % to use more colors
\usepackage{afterpage}
\usepackage{float}
\usepackage{bbold}
\usepackage{hypbmsec}
\usepackage[colorlinks = true,
            linkcolor = blue,
            urlcolor  = blue,
            citecolor = blue,
            anchorcolor = blue]{hyperref}
\allowdisplaybreaks
\usepackage[export]{adjustbox}

\makeatletter

\newcommand{\be}{\begin{equation}}
\newcommand{\ee}{\end{equation}}
\newcommand{\bea}{\begin{eqnarray}}
\newcommand{\eea}{\end{eqnarray}}

\begin{document}

%\title{Leptogenesis via Dark Matter Decay and its imprint on Gravitational Wave}
\title{Baryon asymmetry from dark matter decay in the vicinity of a phase transition}

\author{Debasish Borah}
\email{dborah@iitg.ac.in}
\affiliation{Department of Physics, Indian Institute of Technology Guwahati, Assam 781039, India}

\author{Arnab Dasgupta}
\email{arnabdasgupta@pitt.edu}
\affiliation{Pittsburgh Particle Physics, Astrophysics, and Cosmology Center, Department of Physics and Astronomy, University of Pittsburgh, Pittsburgh, PA 15206, USA}

\author{Matthew Knauss}
\email{mhknauss@wm.edu}
\affiliation{High Energy Theory Group, William \& Mary, Williamsburg, VA 23187, USA}

\author{Indrajit Saha}
\email{s.indrajit@iitg.ac.in}
\affiliation{Department of Physics, Indian Institute of Technology Guwahati, Assam 781039, India}

\begin{abstract}
We propose a novel framework where baryon asymmetry of the universe can arise due to forbidden decay of dark matter (DM) enabled by finite-temperature effects in the vicinity of a first order phase transition (FOPT). In order to implement this novel cogenesis mechanism, we consider the extension of the standard model by one scalar doublet $\eta$, three right handed neutrinos (RHN), all odd under an unbroken $Z_2$ symmetry, popularly referred to as the scotogenic model of radiative neutrino mass. While the lightest RHN $N_1$ is the DM candidate and stable at zero temperature, there arises a temperature window prior to the nucleation temperature of the FOPT assisted by $\eta$, where $N_1$ can decay into $\eta$ and leptons generating a non-zero lepton asymmetry which gets converted into baryon asymmetry subsequently by sphalerons. The requirement of successful cogenesis together with a first order electroweak phase transition not only keep the mass spectrum of new particles in sub-TeV ballpark within reach of collider experiments but also leads to observable stochastic gravitational wave spectrum which can be discovered in planned experiments like LISA. 
\end{abstract}

\maketitle

\noindent
{\bf Introduction:} Presence of dark matter (DM) and baryon asymmetry in the universe (BAU) has been suggested by several astrophysical and cosmological observations \cite{Zyla:2020zbs, Aghanim:2018eyx}. While the standard model (SM) of particle physics fails to solve these two longstanding puzzles, several beyond standard model proposals have been put forward. Among them, the weakly interacting massive particle paradigm of DM \cite{Kolb:1990vq, Jungman:1995df, Bertone:2004pz, Feng:2010gw, Arcadi:2017kky, Roszkowski:2017nbc} and baryogenesis/leptogenesis \cite{Weinberg:1979bt, Kolb:1979qa, Fukugita:1986hr} have been the most widely studied ones. While these frameworks solve the puzzles independently, the similar abundances of DM $(\Omega_{\rm DM})$ and baryon $(\Omega_{\rm B})$ that is, $\Omega_{\rm DM} \approx 5\,\Omega_{\rm B}$ has also led to efforts in finding a common origin or cogenesis mechanism. The popular list of such cogenesis mechanisms include, but not limited to, asymmetric dark matter \cite{Nussinov:1985xr, Davoudiasl:2012uw, Petraki:2013wwa, Zurek:2013wia,DuttaBanik:2020vfr, Barman:2021ost, Cui:2020dly}, baryogenesis from DM annihilation \cite{Yoshimura:1978ex, Barr:1979wb, Baldes:2014gca, Chu:2021qwk, Cui:2011ab, Bernal:2012gv, Bernal:2013bga, Kumar:2013uca, Racker:2014uga, Dasgupta:2016odo, Borah:2018uci, Borah:2019epq, Dasgupta:2019lha, Mahanta:2022gsi}, Affleck-Dine \cite{Affleck:1984fy} cogenesis \cite{Roszkowski:2006kw, Seto:2007ym, Cheung:2011if, vonHarling:2012yn, Borah:2022qln, Borah:2023qag}. Recently, there have also been attempts to generate DM and BAU together via a first order phase transition (FOPT)\footnote{See recent reviews \cite{Mazumdar:2018dfl,Hindmarsh:2020hop, Athron:2023xlk} on FOPT in cosmology.} by utilising the mass-gain mechanism \cite{Baldes:2021vyz}. In \cite{Borah:2022cdx, Borah:2023saq}, a supercooled phase transition was considered where both DM and right handed neutrino (RHN) responsible for leptogenesis acquire masses in a FOPT by crossing the relativistic bubble walls. While the genesis of DM and BAU are aided by a common FOPT in these works, they have separate sources of production. Nevertheless, the advantage of such FOPT related scenarios lies in the complementary detection prospects via stochastic gravitational waves (GW).

In this letter, we propose a novel scenario where DM and BAU have a common source of origin in the vicinity of a FOPT. Though DM is cosmologically stable, it can decay in the early universe due to finite-temperature effects, and could be a viable source of baryon asymmetry. To illustrate the idea, we consider a scenario where a non-zero lepton asymmetry is generated from decay of DM during a short period just before a FOPT and subsequently gets converted into baryon asymmetry. The role of such forbidden decays on DM relic was discussed in several earlier works \cite{Darme:2019wpd, Konar:2021oye,Chakrabarty:2022bcn, Shibuya:2022xkj}. To the best of our knowledge, this is the first time, such forbidden decay of DM facilitated by a FOPT has been considered to be the source of baryon asymmetry of the universe. During a finite epoch in the early universe, just before the nucleation temperature of a FOPT, such forbidden decays of DM, considered to be a gauge singlet RHN, into lepton and a second Higgs doublet is allowed generating a non-zero lepton asymmetry which later gets converted into baryon asymmetry via electroweak sphalerons. The second Higgs doublet not only assists in making the electroweak phase transition (EWPT) first order but also generates light neutrino masses at one-loop level together with the RHNs via the scotogenic mechanism \cite{Tao:1996vb, Ma:2006km}. With all the new fields in sub-TeV ballpark and a strong FOPT, our cogenesis mechanism also has promising detection prospects at particle physics as well as GW experiments.

\vspace{0.5cm}
\noindent 
{\bf The framework:} In order to realise the idea, we consider three RHNs $N_{1,2,3}$ and a new Higgs doublet $\eta$ in addition to the SM particles. Similar to the minimal scotogenic model \cite{Tao:1996vb, Ma:2006km}, these newly introduced fields are odd under an unbroken $Z_2$ symmetry while all SM fields are even. The relevant part of the Lagrangian is given by
\begin{align}
    -\mathcal{L} \supset   \frac{1}{2} M_{ij} \overline{N^c_i} N_j +Y_{\alpha i} \overline{L_\alpha} \tilde{\eta} N_i +{\rm h.c.}
    \label{eq:L}
\end{align}
While neutrinos remain massless at tree level, the $Z_2$-odd particles give rise to one-loop contribution to light neutrino mass \cite{Ma:2006km, Merle:2015ica}. The possibility of FOPT in this model was discussed earlier in \cite{Borah:2020wut,Shibuya:2022xkj}. While \cite{Borah:2020wut} considered single-step FOPT and relevant scalar as well as fermion DM studies, the authors of \cite{Shibuya:2022xkj} studied both single and two-step FOPT and their impact on fermion singlet DM by considering finite-temperature masses. In this work, we assume the FOPT to be single-step for simplicity.

%\begin{align}
%(m_{\nu})_{\alpha \beta} &= \sum_k \frac{Y_{\alpha k}Y_{\beta k} M_{k}}{32 \pi^2} \bigg ( \frac{m^2_{H}}{m^2_{H}-M^2_k} \: \text{ln} \frac{m^2_{H}}{M^2_k} \nonumber \\
%&- \frac{m^2_{A}}{m^2_{A}-M^2_k}\: \text{ln} \frac{m^2_{A}}{M^2_k} \bigg ) 
%\label{numass1}
%\end{align}
%where %$m^2_{R,I}=m^2_{H,A}$ are the masses of scalar and pseudo-scalar part of $\Phi^0_2$ and 
%$M_k$ is the mass eigenvalue of the mass eigenstate $N_k$ assuming the RHN mass matrix to be diagonal. Also, $A, H$ are the neutral pseudoscalar and scalar respectively contained in $\eta$. 

%The thermal, field-dependent masses of different components of $\eta$ can be written in straightforward way. While RHN does not receive much thermal correction to mass, the SM leptons acquire thermal mass as \cite{Giudice:2003jh}
%\begin{align}
%  M_L(T) & = \sqrt{m_L^2+\frac{1}{2}\Pi_\text{gauge}^2(T)},  \nonumber \\
% \Pi_\text{gauge}^2(T) & =\bigg{(}\frac{1}{16}g^{\prime 2}+\frac{3}{16}g^{2}\bigg{)}T^2. 
%\end{align}
%The zero-temperature masses of RHN and $\eta$ components are denoted by $M_i, M_{H, A, \eta^\pm}$ in our discussions.

We calculate the complete potential including the tree level potential $V_{\rm tree}$, one-loop Coleman-Weinberg potential $V_{\rm CW}$\cite{Coleman:1973jx} along with the finite-temperature potential $V_{\rm th}$ \cite{Dolan:1973qd,Quiros:1999jp}. The thermal field-dependent masses of different components of $\eta$ namely neutral scalar $H$, pseudo-scalar $A$, charged scalar $\eta^\pm$ along with other SM particles are incorporated in the full potential. The zero-temperature masses of RHN and $\eta$ components are denoted by $M_i, M_{H, A, \eta^\pm}$ in our discussions. Considering a one-step phase transition, where only the neutral component of the SM Higgs doublet (denoted as $\phi$) acquires a non-zero vacuum expectation value (VEV), we then calculate the critical temperature $T_c$ at which the potential acquires another degenerate minima at $v_c = \phi (T=T_c)$. The order parameter of the FOPT is conventionally defined as $v_c/T_c$ such that a larger $v_c/T_c$ indicates a stronger FOPT. The FOPT proceeds via tunneling, the rate of which is estimated by calculating the bounce action $S_3$ using the prescription in \cite{Linde:1980tt, Adams:1993zs}. The nucleation temperature $T_n$ is then calculated by comparing the tunneling rate with the Hubble expansion rate of the universe $\Gamma (T_n) = \mathcal{H}^4(T_n)$. 

As usual, such FOPT can lead to generation of stochastic gravitational wave background due to bubble collisions~\cite{Turner:1990rc,Kosowsky:1991ua,Kosowsky:1992rz,Kosowsky:1992vn,Turner:1992tz}, the sound wave of the plasma~\cite{Hindmarsh:2013xza,Giblin:2014qia,Hindmarsh:2015qta,Hindmarsh:2017gnf} and the turbulence of the plasma~\cite{Kamionkowski:1993fg,Kosowsky:2001xp,Caprini:2006jb,Gogoberidze:2007an,Caprini:2009yp,Niksa:2018ofa}. The total GW spectrum is then given by 
$$\Omega_{\rm GW}(f) = \Omega_\phi(f) + \Omega_{\rm sw}(f) + \Omega_{\rm turb}(f).$$ While the peak frequency and peak amplitude of such GW spectrum depend upon specific FOPT related parameters, the exact nature of the spectrum is determined by numerical simulations. The two important quantities relevant for GW estimates namely, the duration of the phase transition and the latent heat released are calculated and parametrised in terms of \cite{Caprini:2015zlo}
$$\frac{\beta}{{\mathcal{ H}}(T)} \simeq T\frac{d}{dT} \left(\frac{S_3}{T} \right) $$ and 
$$ \alpha_* =\frac{1}{\rho_{\rm rad}}\left[\Delta V_{\rm tot} - \frac{T}{4} \frac{\partial \Delta V_{\rm tot}}{\partial T}\right]_{T=T_n} $$ 
respectively, where $\Delta V_{\rm tot}$ is the energy difference in true and false vacua. The bubble wall velocity $v_w$ is estimated from the Jouguet velocity \cite{Kamionkowski:1993fg, Steinhardt:1981ct, Espinosa:2010hh}
$$v_J = \frac{1/\sqrt{3} + \sqrt{\alpha^2_* + 2\alpha_*/3}}{1+\alpha_*}$$
according to the prescription outlined in \cite{Lewicki:2021pgr}\footnote{See \cite{Ai:2023see} for a recent model-independent determination of bubble wall velocity.}. We also estimate the reheat temperature $T_{\rm RH}$ after the FOPT due to the release of energy. $T_{\rm RH}$ is defined as $T_{\rm RH} = {\rm Max}[T_n, T_{\rm inf}]$ \cite{Baldes:2021vyz} where $T_{\rm inf}$ is determined by equating density of radiation energy to that of energy released from the FOPT or equivalently $\Delta V_{\rm tot}$. A large reheat temperature can dilute the lepton or baryon asymmetry produced prior to the nucleation temperature by a factor of $(T_n/T_{\rm RH})^3$. Since we are not in the supercooled regime, such entropy dilution is negligible in our case, as we can infer by comparing $T_n, T_{\rm RH}$ for the benchmark points given in table \ref{tab:table1}. In the same table we also show the relevant parameters related to the model and related FOPT, GW phenomenology.

\begin{figure*}
    \centering
    \includegraphics[scale=0.38]{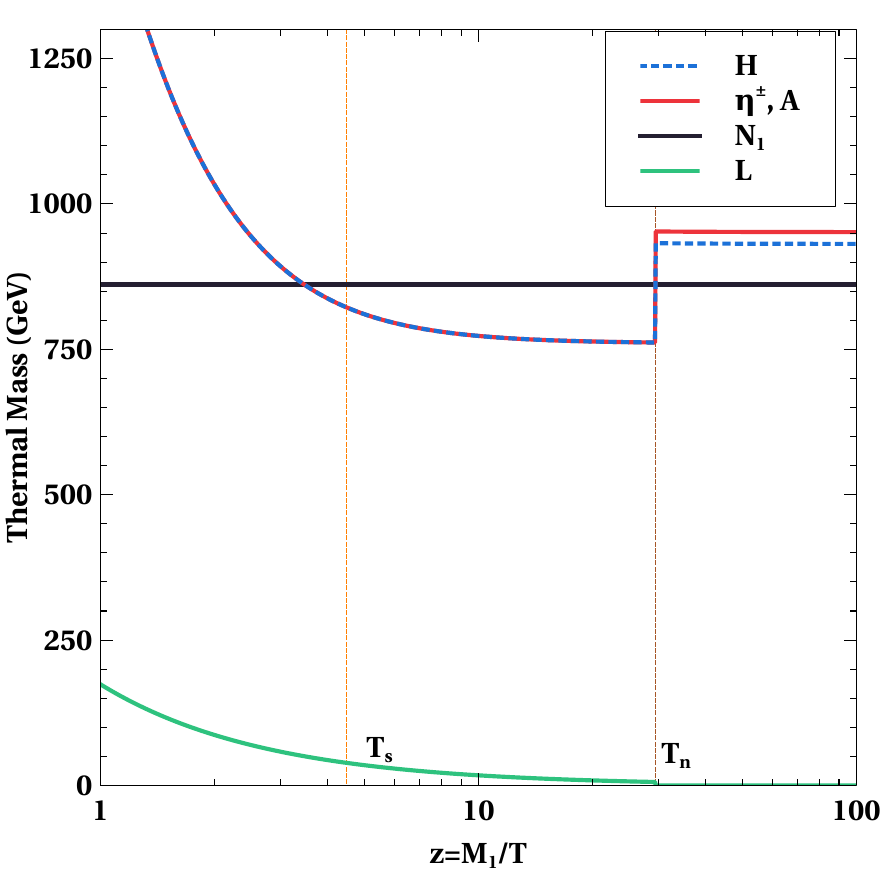}
        \includegraphics[scale=0.38]{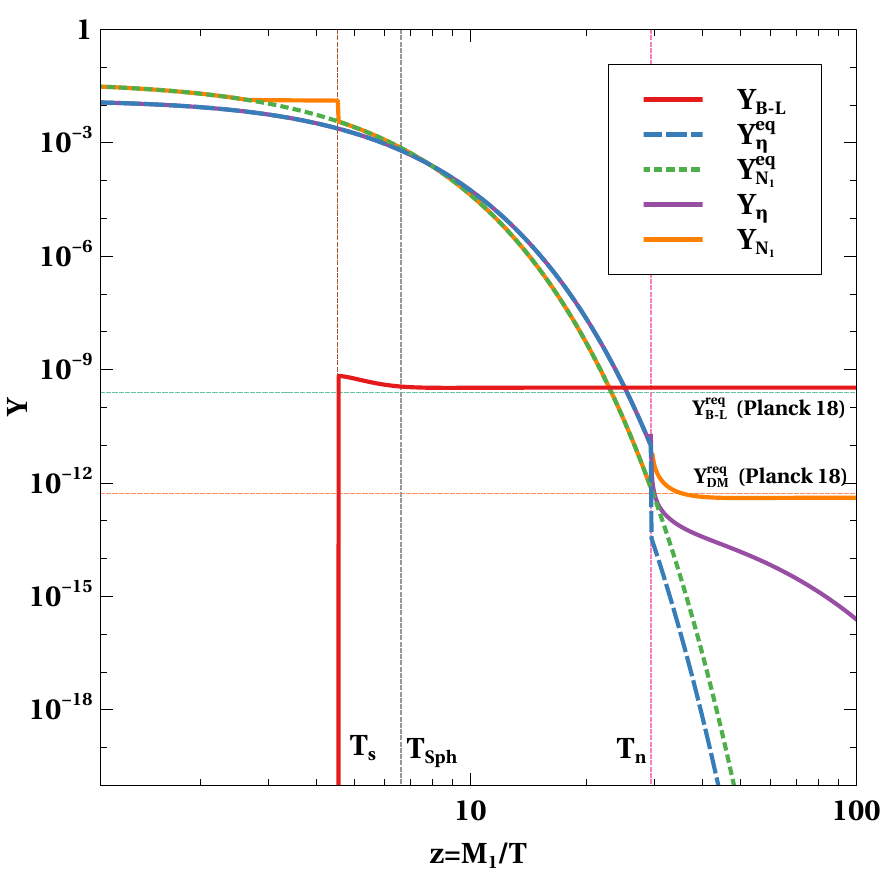}
     \includegraphics[scale=0.38]{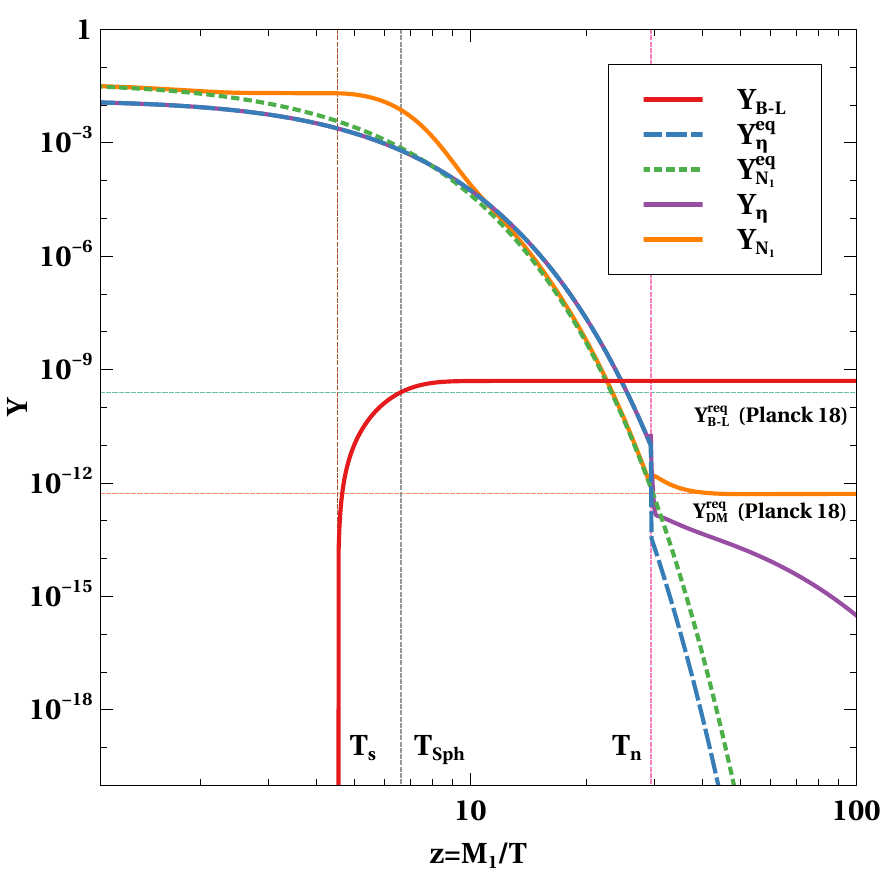} 
    \caption{Left panel: Finite-temperature masses of $L, N_1$ and components of $\eta$ for BP1 shown in table \ref{tab:table1}. Middle panel: Evolution of comoving number densities for $\eta, N_1, B-L$ for BP1 shown in table \ref{tab:table1}(the lightest neutrino mass is $10^{-1}$ eV in normal ordering and $z_{23}=10.34 i$). Right panel: Same as in left panel but for the lightest neutrino mass $10^{-5}$ eV. The vertical line at labelled as $T_s$ ($T_n$) denotes the temperature below which $N_1 \rightarrow L \eta$ decay is kinematically allowed (disallowed). The vertical like labelled as $T_{\rm Sph}$ indicates the sphaleron decoupling temperature of $\sim 130$ GeV.}
    \label{fig:massEv1}
\end{figure*}

\begin{widetext}
\begin{center}
\begin{table}[!h]
    \centering
    \begin{tabular}{|c|c|c|c|c|c|c|c|c|c|c|c|}
    \hline
      & $T_c $ (GeV)   & $v_c$ (GeV) & $T_n$(GeV) & $M_{1}$ (GeV) & $\mu_\eta$ (GeV) & $M_{\eta^\pm}\sim M_A$(GeV) & $M_H$ (GeV) &  $\alpha_*$ & $\beta/{\mathcal{H}}$ & $v_J$ & $T_{\rm RH}$ (GeV) \\
      \hline
       BP1 & 60.05  & 217.22 & 29.27 & 859.50 & 760.25 & 951.51 & 931.26 &  1.29 & 20.21 & 0.94  & 30.37\\
       \hline 
       BP2 & 73.55  & 187.62  & 68.54 & 866.70 & 787.07 & 958.89 & 944.72 & 0.04 & 2862.35 &  0.71 & 68.54\\
       \hline
       BP3  & 71.30 & 199.28 & 64.33 & 676.64 & 579.36 & 774.96 & 743.73 & 0.06 & 1829.84 & 0.74 & 64.33\\
       \hline
       BP4  & 63.35 & 216.65  & 38.49 & 493.74 & 368.04 & 608.38 & 548.60 & 0.45 & 159.33 &  0.88 & 38.49\\
       \hline
    \end{tabular}
    \caption{Benchmark model parameters along with the corresponding FOPT and GW related parameters.Here, $\mu_\eta$ is the bare mass of the inert scalar doublet $\eta$.}
    \label{tab:table1}
\end{table}
\end{center}
\end{widetext}

\vspace{0.5cm}
\noindent
{\bf Cogenesis of baryon and dark matter:} We first discuss the temperature dependence of relevant particle masses leading to the temperature window which enables forbidden decay of DM. The left panel of Fig. \ref{fig:massEv1} shows the temperature dependence of masses of inert scalar doublet components, lepton doublet $L$ and the lightest RHN $N_1$ plotted as a function of $z=M_1/T$ for benchmark point BP1 given in table \ref{tab:table1}. Clearly, $\eta$ remains heavier than $N_1$ at low temperatures, specially after acquiring a new contribution to its mass (in addition to bare mass $\mu_\eta$) from SM Higgs $\Phi$ as a result of the EWPT. This makes $N_1$ the lightest $Z_2$-odd particle at low temperatures and hence cosmologically stable to contribute to DM relic. As seen from the left panel of Fig. \ref{fig:massEv1}, just before the nucleation temperature $T_n$ of EWPT, $\eta$ is lighter than $N_1$, but again becomes heavier at high temperature $T>T_s$ due to large thermal correction. This gives rise to a finite window $(T_n < T < T_s)$ in the vicinity of EWPT where $N_1$ remains heavier than $\eta, L$ enabling the forbidden decay $N_1 \rightarrow \eta L$. Depending upon the duration of this decay and CP asymmetry, it is possible to generate sufficient lepton asymmetry while satisfying DM relic as a result of this forbidden decay. Since we are relying on electroweak sphalerons to convert the lepton asymmetry to baryon asymmetry, we require $T_s > T_{\rm Sph} \sim 130$ GeV. Generation of lepton asymmetry from the lightest RHN decay in minimal scotogenic model was studied in several earlier works \cite{Hambye:2009pw, Racker:2013lua, Clarke:2015hta, Hugle:2018qbw, Borah:2018rca, Mahanta:2019gfe, Mahanta:2019sfo, Kashiwase:2012xd, Kashiwase:2013uy, JyotiDas:2021shia}. Here, we use the finite-temperature corrections which allow $N_1$ to be DM while being responsible for generating lepton asymmetry at high scale, leading to a novel cogenesis possibility in this minimal model.

\begin{figure*}
    \centering
    \includegraphics[scale=0.38]{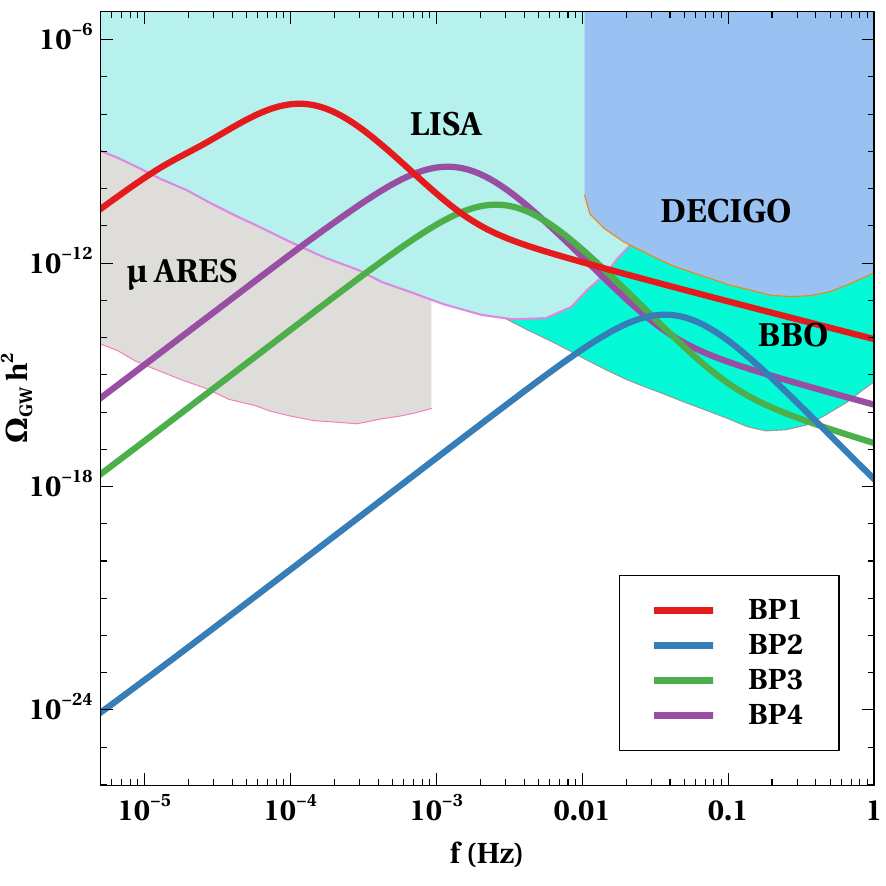}       
    \includegraphics[scale=0.38]{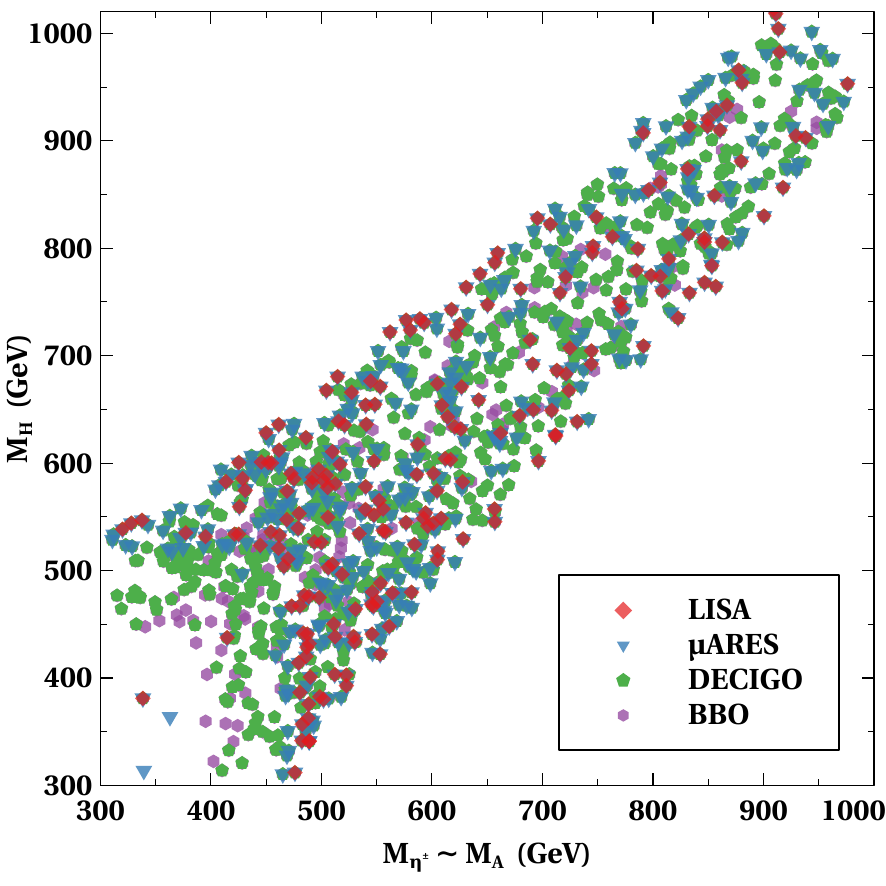}
    \includegraphics[scale=0.38]{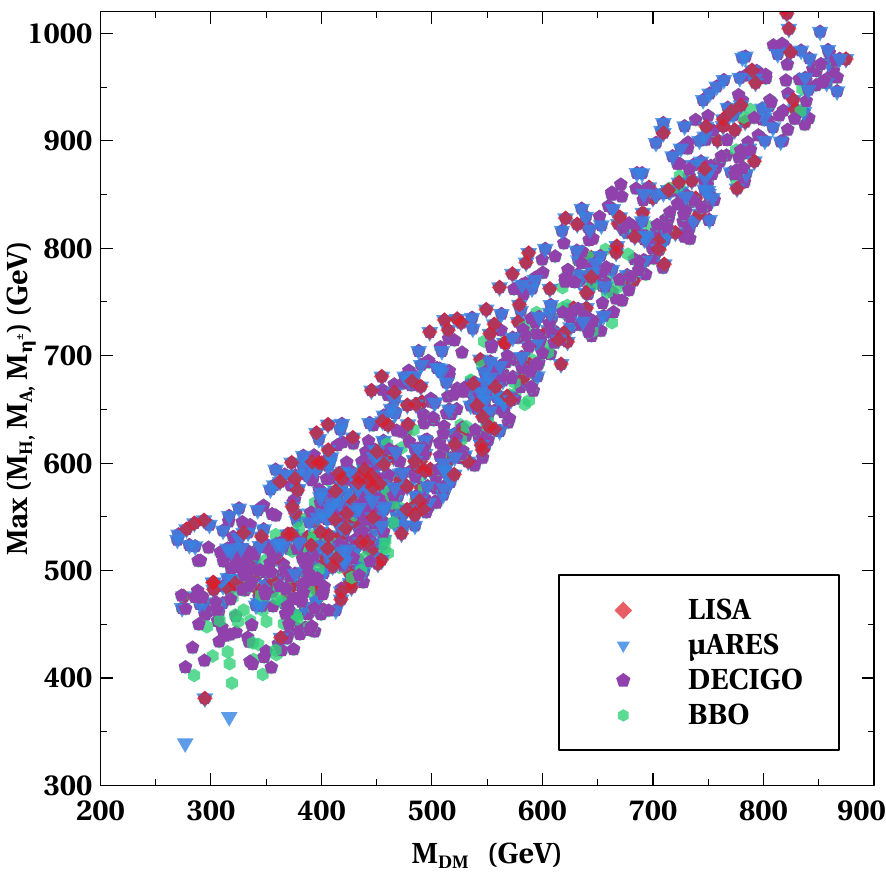}
    \caption{Left panel: GW spectrum corresponding to the benchmark points given in table \ref{tab:table1}. The future sensitivity of LISA, $\mu$ARES, BBO, DECIGO are shown as shaded regions. Middle panel: Parameter space in $M_{\eta^\pm} \sim M_A$ versus $M_H$ plane with colour code showing sensitivity of different GW experiments. Right panel: The parameter space in heaviest Scalar$-M_{\rm DM}$ parameter space with colour code showing sensitivity of different GW experiments. In this scan, $\mu_\eta \in (200-800)$ GeV, $\lambda_2 \in (1, 2)$ , the lightest neutrino mass is $10^{-3}$ eV in NO, $z_{23}=8 i$, the heavier RHN masses are fixed at $M_2=2 M_1, M_3=3 M_1$. The points shown in the scan plots are consistent with DM relic criteria.}
    \label{fig:PTscan1}
\end{figure*}

%\begin{figure}
%    \centering
%    \includegraphics[scale=0.5]{BP_GW.pdf}
%    \caption{GW spectrum corresponding to the benchmark points given in table \ref{tab:table1}. The future sensitivity of LISA, $\mu$ARES, BBO, DECIGO are shown as shaded regions.}
%    \label{fig:GW}
%\end{figure}

In order to find baryon asymmetry and DM relic, the relevant Boltzmann equations for comoving number densities $Y=n_X/s$ of $X \equiv N_1, \eta, B-L$ ($s$ being the entropy density) have to be solved numerically. While we consider self-annihilation of $\eta$ into account in the Boltzmann equations, the (co)annihilation rates for $N_1$ remain much suppressed compared to decay rate of $N_1$ due to small couplings and phase-space suppression. The small Dirac Yukawa couplings of sub-TeV scale $N_1$ are required to satisfy light neutrino masses. The dominant decay and inverse decay rates of $N_1$ are sufficient to keep $N_1$ almost in equilibrium till $T=T_n$. In addition to considering the finite-temperature masses of $N_1, \eta, L$, we also consider the modified CP asymmetry parameter $\epsilon_1$ by appropriately considering such corrections. The lepton asymmetry at the sphaleron decoupling epoch $T_{\rm Sph} \sim 130$ GeV gets converted into baryon asymmetry. The final baryon asymmetry $\eta_B$ can be analytically estimated to be \cite{Buchmuller:2004nz}
 \begin{align}
     \eta_B = \frac{a_{\rm sph}}{f} \epsilon_1 \kappa\,, \label{eqn:etaana}
 \end{align}
where the factor $f$ accounts for the change in the relativistic degrees of freedom from the scale of leptogenesis until recombination and comes out to be $f=\frac{106.75}{3.91}\simeq27.3$. $\kappa$ is known as the efficiency factor which incorporates the effects of washout processes while $a_\text{Sph}$ is the sphaleron conversion factor. The lepton asymmetry at the sphaleron decoupling epoch $T_{\rm Sph} \sim 130$ GeV gets converted into baryon asymmetry as \cite{Harvey:1990qw}
\begin{align}
& Y_B\simeq a_\text{Sph}\,Y_{B-L}=\frac{8\,N_F+4\,N_H}{22\,N_F+13\,N_H}\,Y_{B-L}\,, 
\label{eq:sphaleron}
\end{align}
which, for our model, with $N_F=3\,,N_H=2$ gives $a_\text{Sph}= 8/23$.  

Instead of considering any approximate analytical expressions, we solve the explicit coupled Boltzmann equations involving $N_1, \eta, B-L$ number densities numerically for the same benchmark points shown in table \ref{tab:table1}. The middle and right panels of Fig. \ref{fig:massEv1} show the corresponding evolution of $N_1, \eta$ and $B-L$ for BP1 considering two different values of lightest neutrino mass $m_1$ assuming normal ordering (NO). The heavier RHN masses are fixed at $M_2=2 M_1, M_3=3 M_1$ while the non-zero complex angle in the orthogonal matrix $R$ (which appears in Casas-Ibarra parametrisation \cite{Toma:2013zsa}) is chosen to be $z_{23}=10.34 i$. The quasi-degenerate nature of RHN spectrum is motivated from the fact that the temperature corrected CP asymmetry parameter is derived only for the interference of tree level and self-energy diagrams. For the choices of masses and Casas-Ibarra parameters, Dirac Yukawa couplings of $N_1$ remain at $\lesssim \mathcal{O}(10^{-5})$ while for $N_{2,3}$ they can be as large as $\mathcal{O}(10^{-1})$. Depending upon the lightest neutrino mass $m_1 = 10^{-1}$ eV and $m_1 = 10^{-5}$ eV, leptogenesis can be in strong and weak washout regimes as seen from middle and right panels of Fig. \ref{fig:massEv1} respectively. As clearly seen from both these panels, $Y_{B-L}$ remains zero at $T>T_s$ when $N_1 \rightarrow \eta L$ is kinematically forbidden. Soon after this threshold, lepton asymmetry freezes in and saturates to the asymptotic value at large $z$. After the initial rise in $Y_{B-L}$, the middle panel shows a slight decrease before saturation, typical of a strong washout regime due to larger values of $m_1$ and hence larger Dirac Yukawa couplings associated with $N_1$. For the chosen benchmark satisfying $T_s > T_{\rm Sph} > T_n$, the comoving abundance of RHN $N_1$ saturates at $T<T_n$ giving rise to the required DM relic. While $\eta$ can decay at $T< T_n$, it can not affect baryon asymmetry as $T_n < T_{\rm Sph}$ for BP1. Even for $T_n > T_{\rm Sph}$, $\eta$ decay need not change lepton asymmetry if $\eta \rightarrow \eta^\dagger$ type of processes via scalar portal remains efficient. The late decay of $\eta$ can however, change the abundance of $N_1$. However, for the chosen benchmark point BP1, such late decay contribution to DM abundance is negligible. As seen from the middle and right panels of Fig. \ref{fig:massEv1}, the DM final abundance is consistent with the observed DM relic $\Omega_{\text{DM}} h^2 = 0.120\pm 0.001$ \cite{Aghanim:2018eyx}. Both the strong and weak washout regimes can produce the required lepton asymmetry by $T_{\rm Sph}$ needed to generate observed baryon-to-photon ratio $\eta_B = \frac{n_{B}-n_{\overline{B}}}{n_{\gamma}} \simeq 6.2 \times 10^{-10}$ \cite{Aghanim:2018eyx}. Similar results are also obtained for inverted ordering (IO) of light neutrino masses as well as other choices of benchmark parameters. While we have assumed RHN to be in the bath initially, the generic conclusions do not change even if we consider RHNs to freeze in from the bath. 

%We perform a numerical scan over the inert doublet parameter space and find the region consistent with a FOPT. We in left panel of Fig. \ref{fig:PTscan1}. The right panel of the same figure shows the parameter space in terms of neutral inert scalar mass and DM mass. Some of these points consistent with FOPT are disfavoured as they correspond to $M_1 > M_{H, A, \eta^\pm}$ preventing $N_1$ from being the stable DM candidate. The points ensuring DM stability also satisfy correct relic abundance. The colour code in the left panel plot of Fig. \ref{fig:PTscan1} shows the order parameter of the FOPT while the same on the right panel plot shows the nucleation temperature. Clearly, to have the required cogenesis from forbidden decay of DM $N_1$, the required nucleation temperature of electroweak phase transition of first order needs to be below 34 GeV. In both the plots, we impose the condition $T_s > T_{\rm Sph}$, required to generate lepton asymmetry before sphaleron decoupling.

\vspace{0.5cm}
\noindent
{\bf Detection prospects:} In the left panel of Fig. \ref{fig:PTscan1}, we show the GW spectrum for the benchmark points given in table \ref{tab:table1}. The same table also contains the details of the GW related parameters used for calculating the spectrum. Clearly, the peak frequencies as well as a sizeable part of the spectrum for three benchmark points remain within the sensitivity of planned future experiment like LISA \cite{2017arXiv170200786A}, keeping the discovery prospect of the model very promising. Sensitivities of other future experiments like $\mu$ARES \cite{Sesana:2019vho}, DECIGO \cite{Kawamura:2006up}, BBO\,\cite{Yagi:2011wg} are also shown as shaded regions, covering most part of the GW spectrum for our benchmark points. In order to project the parameter space of the model against GW sensitivities of these experiments, we perform a numerical scan to find the region consistent with a FOPT and DM relic criteria. The parameter space is shown in the middle and right panel plots of Fig. \ref{fig:PTscan1} with the variations in inert doublet scalar and DM masses. In the colour code, we show the reach of different future GW detectors in terms of respective signal-to-noise ratio (SNR) more than 10. The SNR 
is defined as~\cite{Schmitz:2020syl} 
\begin{equation}
\rho = \sqrt{\tau\,\int_{f_\text{min}}^{f_\text{max}}\,df\,\left[\frac{\Omega_\text{GW}(f)\,h^2}{\Omega_\text{expt}(f)\,h^2}\right]^2}\,, 
\end{equation}
with $\tau$ being the observation time for a particular detector, which we consider to be 1 yr. Clearly, all four experiments mentioned above can probe the parameter space. It should be noted that, the allowed parameter space for inert doublet scalars remains within the TeV ballpark in order to have a first order EWPT. This also restricts DM mass in the same ballpark in order to realise the forbidden decay scenario. Note that, all the points in the scan plots shown in Fig. \ref{fig:PTscan1} do not fulfill the criteria for the observed baryon asymmetry. They can however be made to satisfy the required BAU by suitably varying the CI parameter $z_{23}$ without affecting rest of the phenomenology significantly.

Due to the sub-TeV particle spectrum, the model can also have interesting collider prospects due to the inert scalar doublet $\eta$. The model can give rise to same-sign dilepton plus missing energy \cite{Gustafsson:2012aj, Datta:2016nfz}, dijet plus missing energy \cite{Poulose:2016lvz}, tri-lepton plus missing energy \cite{Miao:2010rg} or even mono jet signatures \cite{Belyaev:2016lok, Belyaev:2018ext} in colliders. The model can also have interesting prospects of charged lepton flavour violating decays like $\mu \rightarrow e \gamma, \mu \rightarrow 3e$ due to light $N_1, \eta$ going inside the loop mediating such rare processes. Particularly for fermion singlet DM, such rare decay rates can saturate present experimental bounds \cite{Toma:2013zsa}.

\vspace{0.5cm}
\noindent
{\bf Conclusion:} We have studied a novel way of generating baryon asymmetry and dark matter in the universe from a common source namely forbidden decay of dark matter felicitated by a first order electroweak phase transition. We adopt the minimal scotogenic model to illustrate the idea where an $Z_2$-odd scalar doublet $\eta$ assists in realising a first order EWPT while also leading to the origin of light neutrino mass at one-loop level with the help of three copies of $Z_2$-odd right handed neutrinos. The lightest RHN is the DM candidate and stable at zero temperature. However, finite-temperature effects and dynamics of the FOPT give rise to a small temperature window $T_s > T > T_n$, prior to the nucleation temperature when DM or $N_1$ can decay into $\eta, L$ generating a non-zero lepton asymmetry which can get converted into baryon asymmetry by electroweak sphalerons provided $T_s > T_{\rm Sph}$, the sphaleron decoupling temperature. The DM becomes stable at $T< T_n$ leading to saturation of its comoving abundance at late epochs. The requirement of a first order EWPT, successful cogenesis leading to observed baryon asymmetry and DM relic in this setup forces the mass spectrum of newly introduced particles to lie in sub-TeV range to be probed at collider experiments. On the other hand, the specific predictions for stochastic gravitational wave spectrum can be probed at planned experiments like LISA. Such complementary detection prospects keep the this novel cogenesis setup verifiable in near future. While we considered a single-step FOPT in our work, two-step FOPT can lead to interesting results for cogenesis along with new detection prospects. On the other hand, implementation of this idea to achieve direct baryogenesis at a scale much lower than the sphaleron decoupling temperature can lead to GW with much lower frequencies which can be observed at pulsar timing array (PTA) experiments, and could in fact be a possible explanation for the recent PTA data \cite{NANOGrav:2023gor, Antoniadis:2023ott, Reardon:2023gzh, Xu:2023wog}. We leave such tantalising possibilities to future works. \\

\acknowledgements
The authors thank Marc Sher for useful comments on the work. The work of DB is supported by the Science and Engineering Research Board (SERB), Government of India grant MTR/2022/000575. The work of MK is supported by the National Science Foundation under Grant PHY-2112460. MK thanks Pittsburgh Particle Physics Astrophysics and Cosmology Center (PITT-PACC) at the University of Pittsburgh for their hospitality.

%\bibliographystyle{apsrev}
%\bibliography{ref, refa, refb, ref1, ref0, ref_3body, ref_aks}

\newpage

\begin{widetext}
\centering
\textbf{\Large Baryon asymmetry from dark matter decay in the vicinity of a phase transition: Supplementary Material}
\end{widetext}

\section{First order phase transition}
\label{appen1}
The tree level scalar potential can be written as
\begin{align}
    V_{\rm tree} & =\mu_\Phi^2|\Phi|^2+\mu_\eta^2|\eta|^2+ \lambda_1|\Phi|^4 + \lambda_2|\eta|^4 +\lambda_3 |\Phi|^2|\eta|^2 \nonumber \\
    & +\lambda_4 |\eta^\dagger \Phi|^2 +\lambda_5[(\eta^\dagger \Phi)^2 +{\rm h.c.}]
\end{align}
The scalar fields $\Phi$ and $\eta$ are parameterized as
\begin{align}
    \Phi=\frac{1}{\sqrt{2}} \begin{pmatrix}
        0 \\
        \phi +v
    \end{pmatrix}, \eta= \begin{pmatrix}
        \eta^\pm \\
        \frac{(H +i A)}{\sqrt{2}}
    \end{pmatrix}.
\end{align}
At finite temperature, the effective potential can be written as
\begin{equation}
    V_{\rm eff}= V_{\rm tree}+V_{\rm CW}+V_{\rm th} + V_{\rm daisy}
    \label{eq:Veff}
\end{equation}
The Coleman-Weinberg potential~\cite{Coleman:1973jx} with $\overline{\rm DR}$ regularisation is given by
\begin{align}
V_{\rm CW} = \sum_i (-)^{n_{f}} \frac{n_i}{64\pi^2} m_i^4 (\phi) \left(\log\left(\frac{m_i^2 (\phi)}{\mu^2} \right)-C_i \right),
\end{align}
where suffix $i$ represents particle species, and $n_i,~m_i (\phi)$ are the degrees of freedom (dof) and field dependent masses of $i$'th particle.
In addition, $\mu$ is the renormalisation scale, and $(-)^{n_f}$ is $+1$ for bosons and $-1$ for fermions, respectively. The squared field dependent physical masses with corresponding dof, relevant for the FOPT calculations, are
\begin{widetext}
\begin{align}
     m_{\eta^\pm}^2 (\phi)=\mu_\eta^2 +\frac{\lambda_3}{2}\phi^2 \,\,\, (n_{\eta^\pm} =2, C_{\eta^\pm} =\frac{3}{2}), \,\,\, 
    m_H^2 (\phi)=\mu_\eta^2 +\frac{\lambda_3+\lambda_4 +2\lambda_5}{2}\phi^2 \,\,\, (n_H=1, C_{H} =\frac{3}{2}) \nonumber \\
    m_A^2 (\phi)=\mu_\eta^2 +\frac{\lambda_3+\lambda_4 -2\lambda_5}{2}\phi^2 \, (n_A=1, C_{A} =\frac{3}{2}), \,
    m_W^2(\phi)=\frac{g_2^2}{4}\phi^2 \, (n_W=6, C_{W} =\frac{5}{6}) \nonumber \\
    m_Z^2(\phi)=\frac{g_1^2+g_2^2}{4}\phi^2 \,\,\, (n_Z=3, C_{Z} =\frac{5}{6}), \,\,\,
    m_t^2(\phi)=\frac{y_t^2}{2}\phi^2 \,\,\, (n_t=12, C_{t} =\frac{3}{2}), \,\,\, m_b^2(\phi)=\frac{y_b^2}{2}\phi^2 \,\,\, (n_b=12, C_{b} =\frac{3}{2}).
\end{align}
\end{widetext}
Thermal contributions to the effective potential are given by
\begin{align}
V_{\rm th} = \sum_i \left(\frac{n_{\rm B_i}}{2\pi^2}T^4 J_B \left[\frac{m_{\rm B_i}}{T}\right] - \frac{n_{\rm F_{i}}}{2\pi^2}T^4 J_F \left[\frac{m_{\rm F_{i}}}{T}\right]\right),
\end{align}
where $n_{B_i}$ and $n_{F_i}$ denote the dof of the bosonic and fermionic particles, respectively.
In this expressions, $J_B$ and $J_F$ functions are defined by following functions:
\begin{align}
J_B(x) =\int^\infty_0 dz z^2 \log\left[1-e^{-\sqrt{z^2+x^2}}\right], \label{eq:J_B} \\
J_F(x) =   \int^\infty_0 dz z^2 \log\left[1+e^{-\sqrt{z^2+x^2}}\right].
\end{align}
We also include the Daisy corrections \cite{Fendley:1987ef,Parwani:1991gq,Arnold:1992rz} which improve the perturbative expansion during the FOPT. Out of the two popularly used schemes namely, Parwani method and Arnold-Espinosa method, we use the latter. The Daisy contribution is given by
\begin{equation}
        V_{\rm daisy}(\phi,T) = -\sum_i \frac{g_i T}{12\pi}\left[ m^3_i(\phi,T) - m^3_i(\phi) \right]
\end{equation}
The thermal masses for inert doublet components are $m^2_i(\phi,T)=m^2_i(\phi) + \Pi_S(T)$ while for electroweak vector bosons they are 
\begin{align}
m_{W_L}^2(\phi,T) = m_W^2(\phi) +\Pi_W(T), \nonumber \\
m_{Z_L}^2(\phi,T)=\frac{1}{2}( m_Z^2(\phi) +\Pi_W(T)+ \Pi_Y(T)+\Delta(\phi,T)), \nonumber \\
 m_{\gamma_L}^2(\phi,T)=\frac{1}{2}( m_Z^2(\phi) +\Pi_W(T)+ \Pi_Y(T)-\Delta(\phi,T))
\end{align}
where

\begin{align}
\Pi_S(T) = \bigg (\frac{1}{8}g_2^2+\frac{1}{16}(g_1^2+g_2^2)+\frac{1}{2}\lambda_2+\frac{1}{12}\lambda_3+\frac{1}{24}\lambda_A \nonumber \\ 
 +\frac{1}{24}\lambda_H + \frac{1}{4}y_t^2+\frac{1}{4}y_b^2 \bigg )T^2 \nonumber \\
 \Pi_W(T)=2g_2^2T^2, \,\, \Pi_Y(T)=2g_1^2T^2, \nonumber \\
\lambda_H=\lambda_3+\lambda_4 +2\lambda_5, \,\, \lambda_A=\lambda_3+\lambda_4 -2\lambda_5 
\end{align}
While RHN does not receive much thermal correction to mass, the SM leptons acquire thermal mass as \cite{Giudice:2003jh}
\begin{align}
 M_L(T) & = \sqrt{m_L^2+\frac{1}{2}\Pi_\text{gauge}^2(T)},  \nonumber \\
\Pi_\text{gauge}^2(T) & =\bigg{(}\frac{1}{16}g^{\prime 2}+\frac{3}{16}g^{2}\bigg{)}T^2. 
\end{align}

In order to calculate the bounce action numerically, we use a fit for the actual potential which matches very well with the actual potential. The most generic quartic potential can be written as\cite{Adams:1993zs}
\begin{equation}
    V(\phi) =\lambda \phi^4 -a \phi^3 +b \phi^2 .
\end{equation}
The above potential can be used to calculate the Euclidean action in a semi-analytical approach. We use fitting approach to calculate the action, where we use the temperature dependent effective potential given by Eq. \eqref{eq:Veff} and fitted with the generic quartic potential mentioned above. Following \cite{Adams:1993zs}, we calculate the action in three dimension given as 
\begin{equation}
    S_3=\frac{\pi a}{\lambda^{3/2}}\frac{8 \sqrt{2}}{81}(2-\delta)^{-2}\sqrt{\delta/2} {\beta_1\delta+\beta_2\delta^2+\beta_3\delta^3}
\end{equation}
where $\delta=8\lambda b/a^2$, \, $\beta_1$=8.2938,\, $\beta_2$=-5.5330 and $\beta_3$=0.8180.
\section{Boltzmann equations for cogenesis}
\label{appen2}
The Boltzmann Equations for $N_1, \eta, B-L$ in terms of comoving number density $Y_i=n_i/s$ with $n_i$ being number density of species 'i' and $s=\frac{2\pi^2}{45}g_{*s} T^3$ being entropy density of the universe, can be written as
\begin{widetext}
\begin{equation}
    \begin{aligned}
        \frac{dY_N}{dz} &= \frac{g_\eta \langle \Gamma_\eta \rangle}{z\Tilde{\mathcal{H}}}\left[Y_\eta - \frac{Y_\eta^{eq}Y_N}{Y_N^{\rm eq}}\right] - \frac{g_\eta \langle \Gamma_{N_1} \rangle}{z\Tilde{\mathcal{H}}}\left[Y_N - \frac{Y_N^{\rm eq}Y_\eta}{Y_\eta^{\rm eq}}\right], \\
        \frac{dY_\eta}{dz} &= - \frac{g_\eta \langle\sigma_{\eta\eta}v_{\rm rel}\rangle s}{z\Tilde{\mathcal{H}}}\left[ Y_\eta^2 - {Y_\eta^{\rm eq}}^2 \right] - \frac{g_\eta \langle \Gamma_\eta \rangle}{z\Tilde{\mathcal{H}}}\left[Y_\eta - \frac{Y_\eta^{\rm eq}Y_N}{Y_N^{\rm eq}}\right] + \frac{g_\eta \langle \Gamma_{N_1} \rangle}{z\Tilde{\mathcal{H}}}\left[Y_N - \frac{Y_N^{\rm eq}Y_\eta}{Y_\eta^{\rm eq}}\right], \\
        %\frac{dY_{\Delta L}}{dz} &= \epsilon \frac{g_N \langle \Gamma_N \rangle}{x\Tilde{H}}\left[Y_N - \frac{Y_N^{eq}Y_\eta}{Y_\eta^{eq}}\right] - \frac{Y_{\Delta L}}{Y^{eq}_l}\frac{Y_N^{eq}Y_\eta}{Y_\eta^{eq}}\frac{g_N \langle \Gamma_N \rangle}{x\Tilde{H}} \\
        \frac{dY_{B-L}}{dz} &= -\epsilon_1 \frac{g_\eta \langle \Gamma_{N_1} \rangle}{z\Tilde{\mathcal{H}}}\left[Y_N - \frac{Y_N^{\rm eq}Y_\eta}{Y_\eta^{\rm eq}}\right] - \epsilon_\eta \frac{g_\eta \langle \Gamma_\eta \rangle}{z\Tilde{\mathcal{H}}}\left[Y_\eta - \frac{Y_\eta^{\rm eq}Y_N}{Y_N^{\rm eq}}\right] - \left(W_1 + \Delta W\right)Y_{B-L}.
    \end{aligned}
\end{equation}
\end{widetext}
Here $z=M_1/T$ and $g_\eta=4$ assuming all components of $\eta$ to be degenerate. We also consider $\epsilon_\eta=0$ for simplicity. $\Tilde{\mathcal{H}} \sim \mathcal{H}=\sqrt{\frac{4\pi^3 g_*(T)}{45}}\frac{T^2}{M_{\rm Pl}}$, the Hubble parameter at high temperatures where $g_{*s}$ remains constant. The CP asymmetry parameter corresponding to $N_i \rightarrow \eta L$ decay, while including finite-temperature effects and summing over all lepton flavours, is given by
\begin{widetext}
\begin{equation}
    \epsilon_i =\left[\left(M_{i}^2 + M_{L}^2 - m_\eta^2\right)\lambda^{1/2}\left(M_{i}^2,M_{L}^2,m_\eta^2\right)\Theta\left(M_{i}^2 - \left(m_\eta + M_{L}\right)^2\right)\right]\frac{1}{(Y^{\dagger}Y)_{ii}}\sum_{j\neq i}\frac{{\rm Im}[((Y^{\dagger}Y)_{ij})^{2}]}{16\pi M_{i}^3} \frac{M_{j}\Delta_{ij}}{\Delta_{ij}^2 + \left(M_{j}\Gamma_{N_j}\right)^2} 
\end{equation}
\end{widetext}
where $\Delta_{ij} \equiv M_i^2 - M_j^2, \lambda(x,y,z) \equiv x^2 + y^2 + z^2 - 2xy - 2xz - 2yz $ and $\Theta(x)$ is the Heaviside step function. It should be noted that in the above derivation of temperature corrected CP asymmetry, we consider the interference of tree level and self-energy diagrams only, assuming a quasi-degenerate RHN spectrum for which the vertex diagram contribution remains sub-dominant. In the above Boltzmann equations, $\langle \Gamma_i \rangle, \langle \sigma_{ii} v_{\rm rel} \rangle$ correspond to thermal averaged decay width and self-annihilation cross-section of species 'i' respectively. The washout terms in the Boltzmann equation for $B-L$ are \cite{Hugle:2018qbw}

\begin{align}
%    \begin{aligned}
%        D &\equiv z\frac{\Tilde{\Gamma}}{\Tilde{\mathcal{H}}} \frac{K_1\left(z\right)}{K_2\left(z\right)}\\
        W_1 \equiv \frac{1}{4} z^3  \frac{\Gamma_{N_1}}{\Tilde{\mathcal{H}}(T=M_1)} K_1\left(z\right), \nonumber \\ \Delta W \equiv \frac{36\sqrt{5}M_{\rm Pl}}{\pi^{\frac{1}{2}}g_l\sqrt{g_\ast}v^4} \frac{1}{z^2} \frac{1}{\lambda_5^2} M_1 \bar{m}_\xi^2, \nonumber \\ \bar{m}_\xi^2 \approx 4\xi_1^2m_l^2 + \xi_2^2 m_{h_2}^2 + \xi_3^2 m_h^2,
%    \end{aligned}
\end{align}
where $K_i(z)$'s are the modified Bessel functions of the second kind, $m_{l, h_2, h} $ denoting lightest to heaviest active neutrino masses, $g_l=2$. The parameter $\xi$ is defined below. 

The one-loop neutrino mass is given by \begin{align}
(m_{\nu})_{\alpha \beta} &= \sum_k \frac{Y_{\alpha k}Y_{\beta k} M_{k}}{32 \pi^2} \bigg ( \frac{m^2_{H}}{m^2_{H}-M^2_k} \: \text{ln} \frac{m^2_{H}}{M^2_k} \nonumber \\
&- \frac{m^2_{A}}{m^2_{A}-M^2_k}\: \text{ln} \frac{m^2_{A}}{M^2_k} \bigg ) 
\label{numass1}
\end{align}
where $m^2_{R,I}=m^2_{H,A}$ are the masses of scalar and pseudo-scalar part of $\Phi^0_2$ and 
$M_k$ is the mass eigenvalue of the mass eigenstate $N_k$ assuming the RHN mass matrix to be diagonal. In order to incorporate the constraints from light neutrino masses, we use the 
Casas-Ibarra (CI) parametrisation for radiative seesaw model \cite{Toma:2013zsa} which allows us to write the Yukawa coupling matrix satisfying the neutrino data as
\begin{align}
Y_{\alpha i} \ = \ \left(U D_\nu^{1/2} R^{\dagger} \Lambda^{1/2} \right)_{\alpha i} \, ,
\label{eq:Yuk}
\end{align}
where $R$ is an arbitrary complex orthogonal matrix satisfying $RR^{T}=\mathbb{1}$ and $D_\nu \ = \ U^\dag m_\nu U^* \ = \ \textrm{diag}(m_1,m_2,m_3)$, the diagonal light neutrino mass matrix. The matrix $U$ is the usual Pontecorvo-Maki-Nakagawa-Sakata (PMNS) mixing matrix $U$ which diagonalises the light neutrino mass matrix $m_\nu$ given by Eq. \eqref{numass1} (assuming diagonal charged lepton basis). In Eq. \eqref{eq:Yuk}, the diagonal matrix $\Lambda$ is defined as
\begin{align}
 \Lambda_\alpha \  = \ \frac{2\pi^2}{\lambda_5}\xi_\alpha\frac{2M_\alpha}{v^2}, \,\, 
\textrm {and}\nonumber \\
\xi_\alpha \ = \  \left(\frac{M_{\alpha}^2}{8(m_{\eta_R}^2-m_{\eta_I}^2)}\left[L_\alpha(m_{\eta_R}^2)-L_\alpha(m_{\eta_I}^2) \right]\right)^{-1} \, . \label{eq:zeta}
\end{align}
The loop function $L_k(m^2)$ is defined as 
\begin{align}
L_k(m^2) \ = \ \frac{m^2}{m^2-M^2_k} \: \text{ln} \frac{m^2}{M^2_k} \, .
\label{eq:Lk}
\end{align}
%\end{widetext}

\end{document}